%% file: ms.tex
\documentclass{pasj01}
\usepackage{times}
\usepackage{lscape}
\usepackage{url}
\usepackage{color}
\usepackage{ulem}
\usepackage{longtable}
\usepackage{graphicx}
\usepackage{longtable}
\usepackage{deluxtable}

\newcommand{\kt}{\ensuremath{k_{\rm{B}}T}}
\newcommand{\lx}{\ensuremath{L_{\rm{X}}}}

\newcommand{\fx}{\ensuremath{F_{\rm{X}}}}

\newcommand{\fnir}{\ensuremath{F_{\rm{NIR}}}}

\newcommand{\nh}{\ensuremath{N_{\rm H}}}
\newcommand{\ksi}{\textit{K\ensuremath{_{\rm{s}}}}}

\newcommand{\ergs}{{\rm erg}\ {\rm s}^{-1}} % PASJ
\newcommand{\ergcms}{{\rm erg}\ {\rm cm}^{-2}\ {\rm s}^{-1}} % PASJ
\newcommand{\ergcmsdeg}{{\rm erg}\ {\rm cm}^{-2}\ {\rm s}^{-1}\ {\rm deg}^{-2}} % PASJ

\newcommand{\ji}{\textit{J}}
\newcommand{\hi}{\textit{H}}

\begin{document}
\Received{}
\Accepted{}
\Published{}
\SetRunningHead{Morihana et al.}{Subaru MOIRCS NIR Imaging of faint X-ray point sources in GBXE}
\title{Deep Near-infrared Imaging Observation of the Faint X-ray Point Sources Constituting the Galactic Bulge X-ray Emission}
\author{Kumiko~\textsc{Morihana},\altaffilmark{1,2}
Masahiro~\textsc{Tsujimoto},\altaffilmark{3}
Ken~\textsc{Ebisawa},\altaffilmark{3}
and Poshak Gandhi\altaffilmark{4}
}%
\altaffiltext{1}{Graduate School of Science, Nagoya University, Chikusa-ku, Nagoya, Aichi 464-8602, Japan}
\altaffiltext{2}{Institute of Liberal Arts and Sciences, Nagoya University, Chikusa-ku, Nagoya, Aichi 464-8601, Japan}
\altaffiltext{3}{Japan Aerospace Exploration Agency, Institute of Space and
Astronautical Science,\\3-1-1 Yoshino-dai, Chuo-ku, Sagamihara, Kanagawa 252-5210, Japan}
\altaffiltext{4}{Department of Physics and Astronomy, University of Southampton, Highfield, SO17 1BJ Southampton, UK}

\email{morihana@u.phys.nagoya-u.ac.jp}


\maketitle

\begin{abstract}
 Presence of the apparently extended hard (2--10~keV) X-ray emission along the Galactic
 plane has been known since the early 1980s. With a deep X-ray exposure using the Chandra 
 X-ray Observatory of a slightly off-plane region in the Galactic bulge, most of the extended emission 
 was resolved into faint discrete X-ray sources in the Fe K band \citep{Revnivtsev2009}. The major 
 constituents of these sources have long been considered to be X-ray active stars and magnetic cataclysmic 
 variables (CVs). However, recent works including our NIR imaging and spectroscopic studies \citep{Morihana2013, Morihana2016}
 argue that other populations should be more dominant. To investigate this further, we conducted a much 
 deeper NIR imaging observation at the center of the Chandra's exposure field. We have used the 
 MOIRCS on the Subaru telescope, reaching the limiting magnitude of $\sim$18~mag in the \ji, \hi, and \ksi\
 bands in this crowded region, and identified $\sim$50\% of the X-ray sources with
 NIR candidate counterparts. We classified the X-ray sources into three groups (A, B, and C) based 
 on their positions in the X-ray color-color diagram and characterized them based on the X-ray and NIR features.
 We argue that the major populations of the Group A and C sources are, respectively, CVs (binaries containing  
magnetic or non-magnetic white dwarfs with high accretion rates) and X-ray active stars. The major 
 population of the Group B sources is presumably WD binaries with low mass accretion rates. The Fe 
 K equivalent width in the composite X-ray spectrum of the Group B sources is the largest among the three 
 and comparable to that of the Galactic bulge X-ray emission. This leads us to speculate that there are numerous 
 WD binaries with low mass accretion rates, which are not recognized as CVs, but are the major contributor of the 
 apparently extended X-ray emission.
\end{abstract}
%\linenumbers

%%%%%%%%%%%%%%%%%%%%
\section{Introduction}
%%%%%%%%%%%%%%%%%%%%%
Presence of the apparently extended hard ($\geq$ 2 keV) X-ray emission along the
Galactic plane has been known since the early 1980s (e.g., \cite{Worrall1982}). The
emission has an integrated luminosity of $\sim$1$\times$10$^{38}$ $\ergs$ in the 2--10
keV band (\cite{Koyama1986a,Valinia1998}) with a spectrum described by two-temperature
thermal plasma (\kt $\sim$1 and 5--10 keV). A remarkable feature in the X-ray spectrum
is the strong Fe K emission line complex (e.g., \cite{Koyama1996}, \cite{Ebisawa2008},
\cite{Yamauchi2009}, \cite{Heard2013}) comprised of neutral or low ionization state line
at 6.4 keV (Fe $\mathrm{I}$) and highly-ionized ion lines at 6.7 (Fe $\mathrm{XXV}$) and
7.0 keV (Fe $\mathrm{XXVI}$). In the most recent review \citep{Koyama2018}, three distinct 
components of the Galactic diffuse emission with 
different scale heights are distinguished ; the Galactic center X-ray emission,
Galactic ridge X-ray emission (GRXE), and Galactic bulge X-ray emission (GBXE).
We mainly focus on the GBXE in this paper, with some notes on the GRXE.

With the advent of the Chandra X-ray Observatory \citep{Weisskopf2002}, a large fraction of the GBXE
and GRXE was resolved into faint discrete sources. For the GBXE, \citet{Revnivtsev2009}
conducted a deep observation of a slightly off-plane region in the Galactic bulge
($l=0.^\circ08, b=-1.^\circ42$; the Chandra bulge field or CBF, hereafter) and argued
that $\sim$88\% of the GBXE around the Fe K band was resolved into faint X-ray point sources
down to a flux of $\sim$10$^{-16}$ $\ergcms$ at the 2--10 keV band. For the GRXE,
Ebisawa et al. (\yearcite{Ebisawa2001, Ebisawa2005}) conducted a deep observation on the Galactic plane 
($l=28.^{\circ}5, b=0.^{\circ}0$; the Chandra plane field or CPF, hereafter) and
discussed that there are two distinct classes of the point sources based on the X-ray spectral hardness.

Because of the similarities in the composite X-ray spectra, it is
likely that the same X-ray point source populations contribute to the GBXE and GRXE
with a different fraction. Then, what are the populations of these faint X-ray point
sources? The most popular ideas are X-ray active
stars and magnetic cataclysmic variables (CV) such as intermediate polars (IPs). The
former contributes to the soft emission, while the latter does to the hard emission
(Revnivtsev et al. \yearcite{Revnivtsev2009, Revnivtsev2011},
\cite{Yuasa2012, Hong2012}). 
More recently, importance of other populations is being recognized. For example,
\citet{Nobukawa2016, Yamauchi2016, Xu2016} discussed that non-magnetic CVs should be also a
major population. 

It should be noted that these results are based on X-ray data alone, in which photon
statistics are generally too poor to constrain the nature of individual sources.
Thus, follow-up observations in longer wavelengths are needed. Because of the large
interstellar absorption toward the GBXE and GRXE, near-infrared (NIR) observations are more
suited than such optical observations that can access CVs only within $\sim$2 kpc
\citep{Motch2010}. In our previous work \citep{Morihana2012}, we conducted NIR imaging
observations of the CBF using the Simultaneous Infrared Imager for Unbiased Survey
(SIRIUS: \cite{Nagayama2003}) on the InfraRed Survey Facility (IRSF) telescope in
the South African Astronomical Observatory. With a limiting magnitude of 16~mag in the
\ji, \hi, and \ksi-band, we identified candidate NIR counterparts to $\sim$11\% of the X-ray point
sources. Upon these NIR-identified X-ray sources, we further carried out NIR
spectroscopic observations using the SofI~\citep{Moorwood1998} on the New
Technology Telescope and the MOIRCS (Multi-Object InfraRed Camera and Spectrograph; \cite{Ichikawa2006, Suzuki2008}) 
on the Subaru in the
multi-object spectroscopy mode~\citep{Morihana2016}. In most soft X-ray sources, we
found NIR spectra with absorption features, which represent X-ray active stars. In only a few
hard X-ray sources, we found NIR spectra with H and He emission lines, which represent
CVs. Unexpectedly, we found 
a dozen of hard X-ray sources exhibiting NIR spectra with
absorption features. We argued that they are new populations different from X-ray active
stars or CVs. In fact, the new population is the most dominant to account for the Fe K
emission line of the GBXE (Morihana et al.~\yearcite{Morihana2013, Morihana2016}). However, due to the
shallow limiting magnitude, the NIR identification rate was low, which did not allow us
to accumulate sufficient samples to characterize and unveil the nature of these
populations.

Here, we present the results of the NIR imaging survey of the CBF using the 8.2~m
Subaru telescope to achieve much deeper observation than our previous study using the 1.4~m IRSF telescope. 
The outline of this paper is as follows. We present the observation and data set in \S~\ref{s2} and data
reduction, analysis, and grouping in \S~\ref{s3}.  In \S~\ref{s4}, we characterize each group of the point sources based
on their X-ray and NIR characteristics, and speculate of the nature of the major population for
each group. The summary is given in \S~\ref{s5}.

%%%%%%%%%%%%%%%%%%%%%%%%%%%%%%%%%%%%%%%%
\section{Observations}\label{s2}
%%%%%%%%%%%%%%%%%%%%%%%%%%%%%%%%%%%%%%%%
A deep \ji\hi\ksi~imaging observation of the CBF was carried out using the MOIRCS
on the
Subaru telescope. MOIRCS is equipped with two 2048 $\times$ 2048 HgCdTe HAWAII-2 arrays
and covers a 4$\times$7 arcmin$^{2}$ field of view with a pixel scale of 0\farcs117
pixel$^{-1}$ in the imaging mode.

\begin{figure}[htbp]
 \begin{center}
  \vspace{-0.5cm}
  \includegraphics[width=0.45\textwidth]{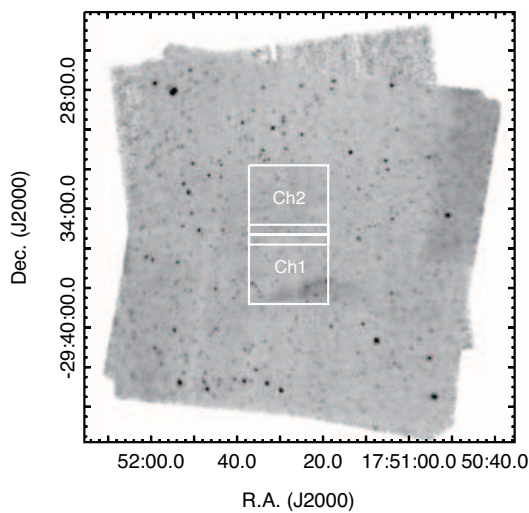}
 \end{center}
 \caption{Subaru/MOIRCS field layout superposed on the Chandra X-ray image in the CBF
 \citep{Revnivtsev2009}. The two white boxes represent the views of the two MOIRCS
 detector chips.}\label{f1}
\end{figure}

We focused on the center region in the CBF, in which the X-ray position determination
accuracy is the best. Among 2002 X-ray point sources detected down to $\sim$10$^{-16}$
$\ergcms$ (2--8 keV; \cite{Revnivtsev2009}) in the CBF, $\sim$540 sources are located
within our MOIRCS field of view \citep{Morihana2013}. Figure~\ref{f1} shows the field
layout of the MOIRCS observations.
The observation was conducted on 2012 May 8 with a seeing of 0\farcs38--0\farcs70. We
used the standard circular dithering pattern of the MOIRCS imaging mode (one center and
eight surrounding pointing) with a dithering amplitude of $\sim$15\arcsec~. Each frame
was exposed for 21 s (\ji~and \ksi) and 16 s (\hi) to avoid saturation of the
detectors. The total exposure time was $\sim$25 (\ji), $\sim$30 (\hi), and $\sim$25
minutes (\ksi).

Because the observed field is very crowded, we could not estimate the
sky background level from the same image. We thus retrieved the MOIRCS archive data of
non-crowded regions with similar observation setup and as close in time with our
observation for each band for assessing the sky level. They are tabulated in
Table\ref{t01}.

\input{t01}

%%%%%%%%%%%%%%%%%%%%%%%%%%%%%%%%%%%%%%
\section{Data reduction \& analysis}\label{s3}
%%%%%%%%%%%%%%%%%%%%%%%%%%%%%%%%%%%%%%
\subsection{Reduction and source extraction}\label{s3-1}
%%%%%%%%%%%%%%%%%%%%%%%%%%%%%%%
We used the \texttt{MCSRED} pipeline\footnote{See
\url{https://www.naoj.org/staff/ichi/MCSRED/mcsred.html} for detail.} developed for
MOIRCS imaging data reduction based on the Infrared Reduction Analysis Facility
(IRAF\footnote{See \url{http://iraf.noao.edu/} for detail.}). It executes dark frame
subtraction, flat fielding, sky subtraction, correction for the optical distortion,
measurements of positional offsets among dithered frames, and stacking the frames into
one. The reduction was performed independently for each chip and band. They were
combined into one image per band in the end.

We searched for sources in the \ji, \hi, and \ksi~bands separately using the
\texttt{SExtractor} software (\cite{Bertinl1996}). For the source extraction, we used a
Gaussian smoothing filter with an FWHM of 3 pixels over 3$\times$3 pixels. We detected
23,328,  23,267,  and 21,903 sources in the \ji-, \hi-, and \ksi-bands down to $\sim$18
mag, respectively.

In order to calibrate the astrometry and photometry of the MOIRCS sources, we consulted
to other NIR source catalogs in this region. In our previous work \citep{Morihana2012},
we presented the results of NIR imaging of the entire CBF using the 2MASS
\citep{Skrutskie2006} and the SIRIUS data. However, we could not utilize these catalogs
because the most 2MASS and SIRIUS sources are saturated in the MOIRCS image. Just in between
the dynamic range gap, there exist VISTA (the Visible and Infrared Survey Telescope for
Astronomy) Variable in the Via Lactea (VVV) near-infrared imaging survey data release 2
\citep{Minniti2018}. The survey depth of the VISTA data is \ksi$\sim$16.5 mag
\citep{Saito2012}, which is at least a magnitude deeper than those of 2MASS and SIRIUS.

\begin{figure*}[htbp]
 \begin{center}
 \includegraphics[width=0.95\textwidth]{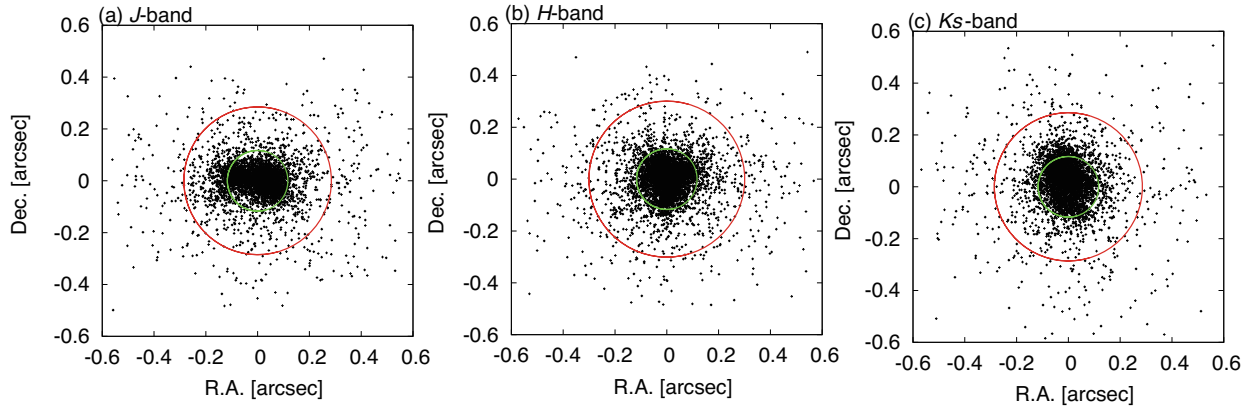}
   \hspace{-10cm}
 \end{center}
 % \vspace{-1cm}
 \caption{Displacements of MOIRCS from VISTA sources after the astrometric correction
 separately for the \ji, \hi, and \ksi-band. The red and green circles show 3$\sigma$
 radius of the displacement and 1 pixel (0\farcs117) radius of MOIRCS, respectively.}
 \label{f2}
\end{figure*}

\begin{figure*}[htbp]
 \begin{center}
  % \vspace{-2cm}
  \begin{tabular}{c}
   \begin{minipage}{0.32\hsize}
    \begin{center}
    \includegraphics[width=0.95\textwidth]{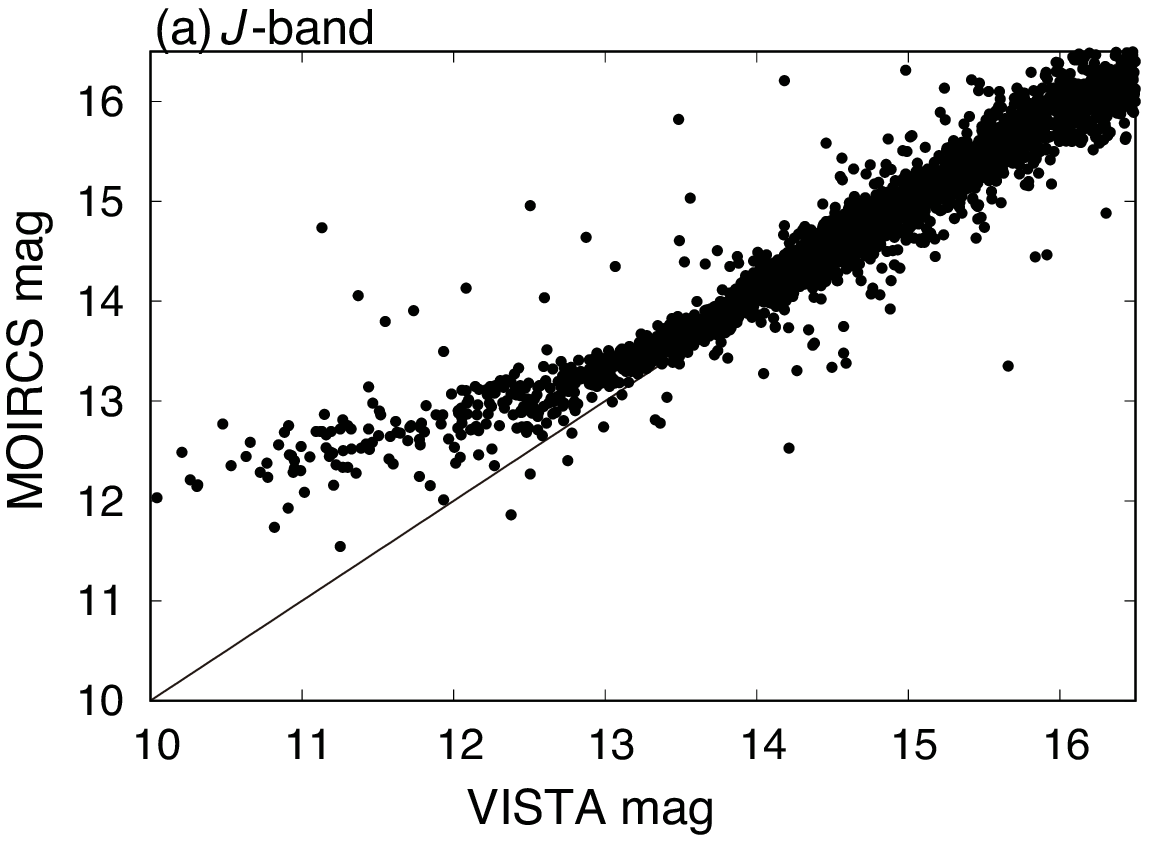}
     \hspace{-3cm}
    \end{center}
   \end{minipage}

   \begin{minipage}{0.32\hsize}
    \begin{center}
     \includegraphics[width=0.95\textwidth]{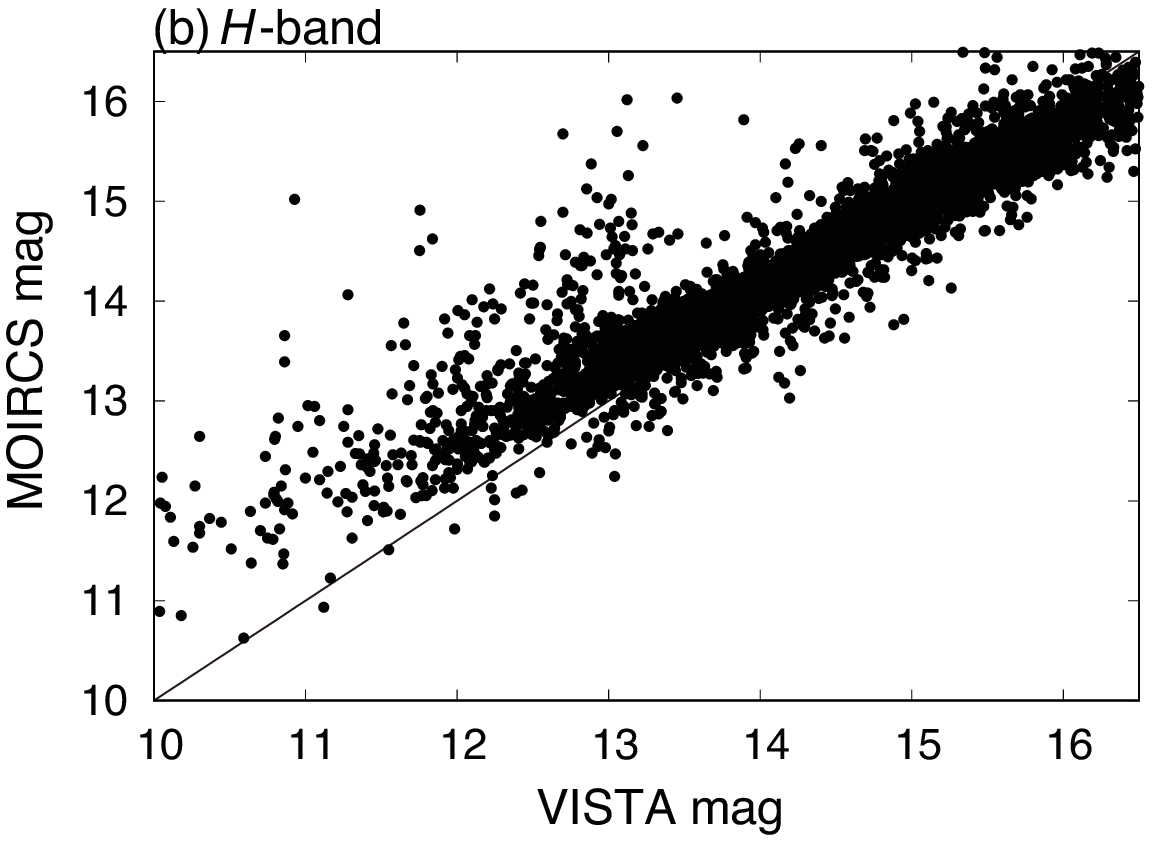}
     \hspace{-3cm}
    \end{center}
   \end{minipage}

   \begin{minipage}{0.32\hsize}
    \begin{center}
     \includegraphics[width=0.95\textwidth]{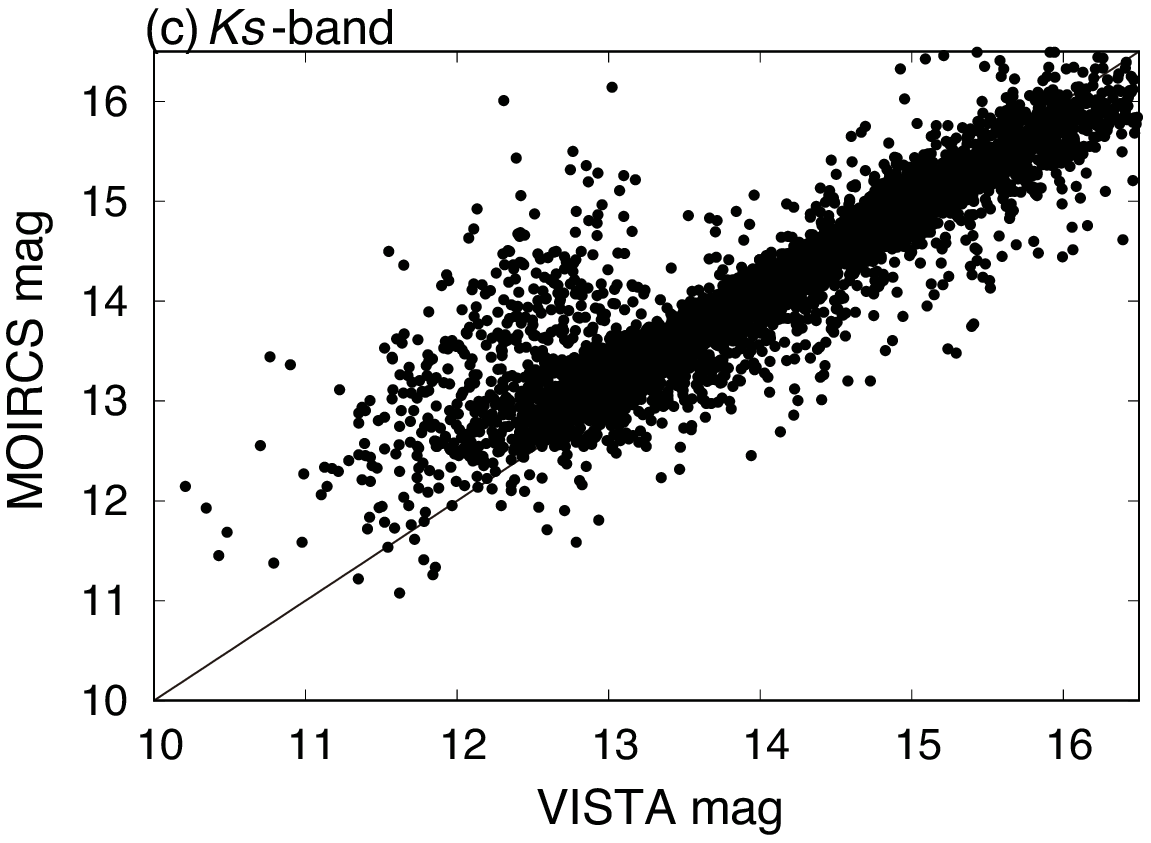}
     \hspace{-1.6cm}
    \end{center}
   \end{minipage}

  \end{tabular}
 \end{center}
 \vspace{-0.2cm}
 \caption{Comparison of MOIRCS and VISTA photometry separately for the \ji, \hi, and
 \ksi-band.}\label{f3}
\end{figure*}

We cross-correlated the MOIRCS and VVV sources and corrected the astrometry and
photometry of the MOIRCS sources compared with the VVV sources. In
figure~\ref{f2}, we compare VISTA and MOIRCS positional differences after the astrometric
correction. The standard deviation of the systematic displacement between the VISTA and 
MOIRCS sources is 0\farcs078 in R.\,A.  and 0\farcs081 in Dec., which are smaller than the 
pixel scale of MOIRCS (0\farcs117). 
After the correction of the MORICS positions to match with 
those of VVV, the residual offset is $<$0\farcs002 in both R.\,A. and Dec. directions in all bands.
In figure~\ref{f3}, we show correlations between VISTA and MOIRCS \ji \hi \ksi~magnitudes
after the photometric correction. MOIRCS magnitudes are fainter than those of VISTA for
the sources brighter than 13.5 mag, which is due to the saturation in MOIRCS. In the
photometric calibration, therefore, we used only the sources in the magnitude ranges of 13.5--16.5
mag (\ji), 13.5--16.5 mag (\hi), and 13.5--16.5 mag (\ksi).

To estimate the completeness of our data, we embedded 200 artificial objects in the
MOIRCS image and attempted detecting them using the same source extraction
algorithm. 
Here, we used the \texttt{mkobject} task of IRAF to embed 200 artificial
sources in the magnitude ranges from 14.0 to 18.0 mag with a 0.5 mag step. We derived the 
detection rates for each range using the same detection algorithm. 
We regarded as detection if the embedded sources were detected at the same position
with the same magnitude range as the input values within a margin.
We estimated the 90\% and 50\% completeness limits for each band  as follows;  17.0 mag and 
18.0 mag in \ji~band, 16.5 mag and 17.5 mag in \hi~band, and 16.5 mag and 17.5 mag in \ksi~band.

%%%%%%%%%%%%%%%%%%%%%%%%%%%%%%%%%%%%%%%%%%%%%%%%%%%%%%%%%%%%%%
\subsection{NIR identification of X-ray sources}\label{s3-2}
\begin{figure}[h]
 \begin{center}
  \includegraphics[width=0.5\textwidth]{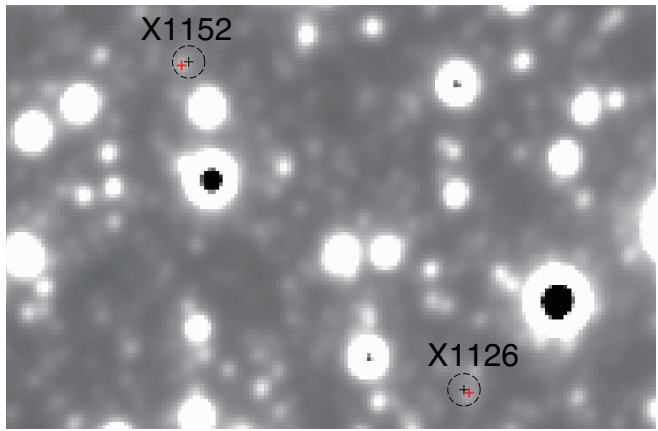}
  \hspace{-10cm}
 \end{center}
 % \vspace{-1cm}
 \caption{Examples of the new MOIRCS identification of the X-ray sources. The X-ray source
 positions (source names starting with X in \cite{Morihana2013}) are shown with
 black crosses and their typical positional uncertainty (0.5\arcsec~) with broken circles on the
 MOIRCS \ksi-band image. 
 Their NIR counterpart candidates are shown with red crosses.
 \ksi-band magnitude of the two sources are 16.88 mag (X1126) and 16.21 mag (X1152), respectively.}
 \label{f4}
\end{figure}

%%%%%%%%%%%%%%%%%%%%%%%%%%%%%%%%%%%%%%%%%%%%%%%%%%%%%%%%%%%%%%
We look for candidate NIR counterparts to the Chandra X-ray sources.
For the NIR sources brighter than 13.5 (\ji), 14.0
(\hi), and 14.0 (\ksi) magnitudes, we use the result of our previous work
\citep{Morihana2012}. In the present work, we extend the identification to fainter NIR
magnitudes using the MOIRCS data. 
The search radius was set as 0\farcs5, which is comparable to the quadrature sum (0\farcs62) of the 
relative and absolute astrometric errors of X-ray and NIR sources.
The following known terms were added in quadrature : 
(a) the error in the absolute astrometry of Chandra (0\farcs58; ACIS-I, \footnote{See \url{https://cxc.harvard.edu/cal/ASPECT/celmon/}}),
(b) the relative astrometry errors of Chandra sources (values in table 3, we adopted a typical value of 0\farcs2),
(c) the error in the absolute astrometry of VISTA (0\farcs07, \cite{Saito2012}), and
(d) the relative astrometry errors of the NIR sources (0\farcs112; figure~\ref{f2}).
First, we searched for the nearest NIR source for each X-ray source and also the nearest X-ray source
for each NIR source. We then regarded them candidate NIR counterparts if a NIR and X-ray source pair is closest to each other. 
An example is shown in figure~\ref{f4}. 
The resultant offset between the X-ray and NIR
sources is 0\farcs085, which is sufficiently small in comparison with the search radius.
As a result, we
identified MOIRCS counterparts to $\sim$40\% of the X-ray sources. This is significantly
larger than our previous result of $\sim$11\% using 2MASS and SIRIUS surveys
\citep{Morihana2012}. The statistics are shown in table~\ref{t02} and the counterpart table
is shown in table~\ref{t03}. Both tables include the results of \citet{Morihana2012} and
the present study. All together, we identified $\sim$50\% of the X-ray sources with candidate NIR counterparts.

\input{t02}

\input{t03}

In the crowded regions such as the one presented here, false positive counterparts
(unrelated pairs identified as counterparts) are unavoidable. We assessed the number of
the false positive counterparts ($N_\mathrm{FP}$) by
\begin{center}
 \begin{equation}
  N_{\mathrm{FP}} = N_{\mathrm{X}} \pi r^{2}\Sigma,
 \end{equation}
\end{center}
where $N_{\mathrm{X}}$ is the number of the X-ray sources, $r=$ 0\farcs5 is the counterpart
search radius, and $\Sigma$ is the surface number density of NIR sources, which varies as a 
function of the magnitude in each band (table \ref{t04}). The false positive rate increases as the 
magnitude increases in general. 
As shown in table \ref{t04}, 
the false positive is estimated to be more than a half of the candidates
in the range of 17--18 mag in \hi- and \ksi-bands. Thus, we discard the candidates with \hi-
or \ksi-band magnitudes larger than 17~mag in the following analysis. Except for
these, the average false positive rate is 25\%.
To validate the false positive number estimates, we did the following exercise. 
We artificially displaced the relative positions of the X-ray and NIR sources in the R.\,A. or Declination 
directions by $\pm$1--10\arcsec~ and identified the counterpart pairs using the same method. The 
resultant number of counterpart pairs at large displacements are found consistent with the number of
false positive estimates in table \ref{t04}.

\input{t04}

We also assessed the number of false negatives
($N_{\mathrm{FN}}$; related pairs not identified as counterparts due to the separation of X-ray and NIR sources    
    are larger than search radius) by
\begin{center}
 \begin{equation}
  N_{\mathrm{FN}} = N_{\mathrm{X}} f e^{-\left(\frac{r}{\sigma}\right)^{2}},
 \end{equation}
\end{center}
where $f$ is the ratio of the X-ray sources with NIR counterparts (to all X-ray sources
and $\sigma=$ 0\farcs16 is the quadrature sum of the typical X-ray (table~\ref{t03}) and NIR (figure~\ref{f2})
positional uncertainty. We found that the false negative is negligible (less than 1\% of the number of the false positive) in all the
   magnitude ranges.

%%%%%%%%%%%%%%%%%%%%%%%%%%%%%%%%%%%%%
\subsection{Grouping}\label{s3-3}
\begin{figure}
 \begin{center}
  \includegraphics[width=0.45\textwidth]{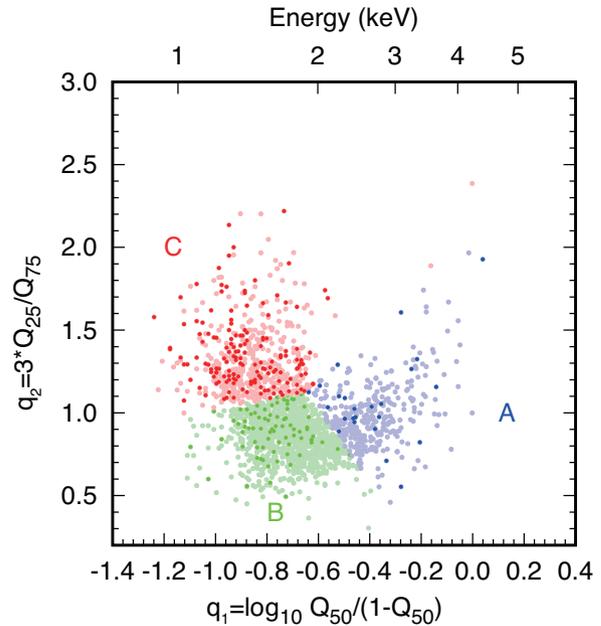}
  \hspace{-10cm}
 \end{center}
 \caption{Quantile distribution of all the X-ray sources detected in the CBF (pale colors;
 \cite{Morihana2013}) and those with NIR counterparts (dark colors). Different
 colors are used for the three groups. The median energy equivalent to each $q_1$ is
 given in the upper horizontal axis.}
 \label{f5}
\end{figure}

%%%%%%%%%%%%%%%%%%%%%%%%%%%%%%%%%%%%
For all the X-ray sources with NIR counterparts, we classify them into groups based on
their X-ray colors. Most of the X-ray sources have poor photon statistics, thus we used the
method based on the color quantiles developed by \citet{Hong2004}, which was also adopted in our
previous work \citep{Morihana2013}. Here, the quantile $E_{x}$ (keV) is the energy
below which $x$\% of X-ray photons reside in the energy-sorted list of photons. The
quantile is normalized as
\begin{equation}
 Q_{x}=\frac{E_{\rm{x}}-E_{\rm{min}}}{E_{\rm{max}}-E_{\rm{min}}},
\end{equation}
in which $E_{\rm{min}}$ and $E_{\rm{max}}$ are the lower and upper bounds of the energy
(0.5 and 8~keV), respectively. We used $Q_{25}$, $Q_{50}$, and $Q_{75}$, and defined two
metrics as $q_1 = \log_{10}
Q_{\rm{50}}/\left(1-Q_{\rm{50}}\right)$ and $q_2 = 3Q_{\rm{25}} / Q_{\rm{75}}$. Here,
the metric $q_1$ reflects the degree of photon spectrum being biased toward the harder
($q_1$ is large) or softer ($q_1$ is small) end, and the metric $q_2$ reflects the
degree of photon spectrum being less ($q_2$ is large) or more ($q_2$ is small)
concentrated around the peak. The metric $q_1$ is a proxy to the more often-used hardness
ratio or the median energy. We plotted all the X-ray sources detected in the CBF  in the
$q_1$-$q_2$ plane (figure~\ref{f5}).  
The distribution is characterized by a concentration at $q_1 \sim -0.7$ and $q_2 \sim
0.8$ and two branches extending upward and rightward. Following our previous work
\citep{Morihana2013}, we used the $k$-means clustering algorithm method \citep{Macqueen67}, which defines
groups to minimize the distance from each source to the centroid of each group. As a
result, we qualitatively classified the sources into three groups A, B, and C.
We estimate the false positive rate of each group to be 41\% (A), 34\% (B), and 25\% (C) from the expected false 
positive numbers derived from large search radii.

%%%%%%%%%%%%%%%%%%%%%%%%%%%%%%%%%%%
\section{Discussion}\label{s4}
%%%%%%%%%%%%%%%%%%%%%%%%%%%%%%%%%%
We now discuss nature of the faint X-ray sources based on the grouping defined in \S
\ref{s3-3}. Given the rate of false positive NIR counterparts (\S~\ref{s3-2}), we limit ourselves to
discuss the overall characteristics of groups and not individual sources.
In \S~\ref{s4-1}, we derive general characteristics of each group as follows; NIR 
and optical identification rates (\S~\ref{s4-1-1}), composite X-ray spectrum
(\S~\ref{s4-1-2}), X-ray variability (\S~\ref{s4-1-3}), X-ray to NIR flux ratio
(\S~\ref{s4-1-4}), and the distance and X-ray luminosity based on the parallax measurements
by Gaia (\S~\ref{s4-1-5}). In \S~\ref{s4-2}, we discuss the likely nature for the
primary constituent of each group. In \S~\ref{s4-3}, we discuss the populations that
mostly contribute to the GBXE.
%

%%%%%%%%%%%%%%%%%%%%%%%%%%%%%%%%%%%%%%%%%%%%%%%%%%%%%%%%%%%%%%%%
\subsection{Characteristics of the X-ray sources}\label{s4-1}
%%%%%%%%%%%%%%%%%%%%%%%%%%%%%%%%%%%%%%%%%%%%%%%%%%%%%%%%%%%%%%%%%
%%%%%%%%%%%%%%%%%%%%%%%%%%%%%%%%%%%%%%%%%%%%%%%%%%%%%%%%
\subsubsection{NIR and optical identification rates}\label{s4-1-1}
%%%%%%%%%%%%%%%%%%%%%%%%%%%%%%%%%%%%%%%%%%%%%%%%%%%%%%%%
Table~\ref{t05} shows a summary of our NIR identification rate of the X-ray sources
for each group. In addition, we retrieved the Gaia data release 2 \citep{Prusti2016}.
%with which we can obtain the parallax distance and hence luminosities of some of the X-ray sources. 
We searched for  Gaia \textit{G}-band
(0.33--1.0 $\mu$m) counterparts of all the X-ray sources using the same method as in
\S~\ref{s3-2}. With a search radius of 0\farcs5, we identified the Gaia counterparts to
109 X-ray sources down to 19.5~mag. 
Among the 109 sources, six of them were identified using optical images taken with the Hubble Space Telescope (tables
   3--5 in \cite{Verg2009}).
The false positive rate is
estimated to be $\sim$17\%. About a half of them have a distance estimate based on
the parallax. The optical identification rate is also given in table~\ref{t05}.

\input{t05}

%%%%%%%%%%%%%%%%%%%%%%%%%%%%%%%%%%%%%%%%%%%%%%%%%%%%%%%%
\subsubsection{X-ray composite spectra}\label{s4-1-2}
%%%%%%%%%%%%%%%%%%%%%%%%%%%%%%%%%%%%%%%%%%%%%%%%%%%%%%%%
\begin{figure*}[htbp]
 \begin{center}
  \includegraphics[width=0.95\textwidth]{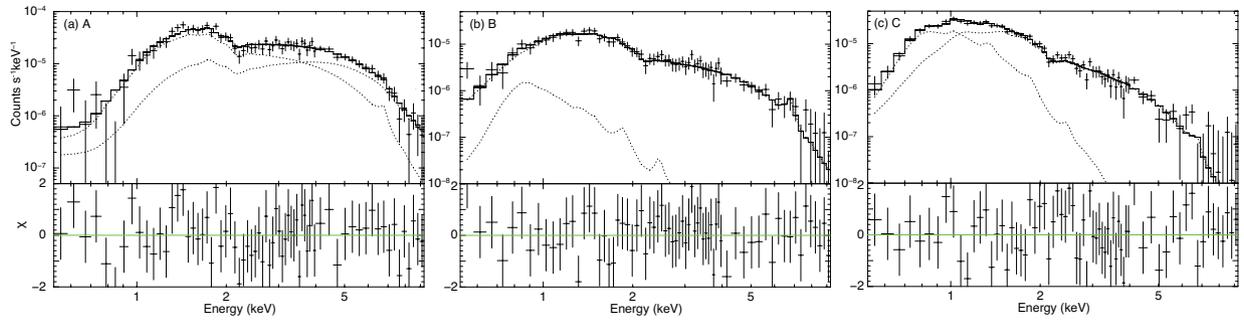}
  \hspace{-10cm}
 \end{center}
 \caption{Composite X-ray spectra and the best-fit models of the three groups (A, B, and
 C). Grouped data (pluses) and the best-fit model convolved with the instrumental
 response (solid histograms) are shown in the upper panel, while the residuals to the
 fit are shown in the lower panel. The best-fit parameters are given in table~\ref{t06}.}
 \label{f6}
\end{figure*}

We made a composite 0.5--8~keV spectrum of all the X-ray sources in each group. We used
the \texttt{combine\_spectra} tool in the \texttt{CIAO} package, which makes averaged combined
background-subtracted spectra as well as count-weighted telescope and instrument
response functions. We binned the spectra to have more than 20
counts per bin and fitted them based on the $\chi^{2}$ statistics.

\input{t06}

For the spectral models, we used 
optically-thin thermal plasma model \citep{Smith2001} or a power-law model attenuated by the
interstellar photo-electric absorption model \citep{Wilms2000}. Free parameters for the
thermal model are the plasma temperature (\kt), the metal abundance ($Z$), and the flux
(\fx) in the 0.5--8.0 keV band.
Those for the power-law model are the photon index ($\Gamma$) and the
flux (\fx). Free parameter for the absorption model is the absorption column density
(\nh). We regarded the fitting to be unsuccessful if the reduced $\chi^{2}$ was larger
than 1.5 or the best-fit parameters were unphysical; $\Gamma>3$ in the power-law
fitting or \kt$>15$~keV in the thermal fitting. 

We started fitting the spectra with a single thermal plasma or a single power-law
model. The fitting was not successful for all the groups. We then added a second continuum
component and fitted the spectra with two thermal components of different temperatures
or a one-temperature thermal plus a power-law component. The fitting was successful for all
the groups. When both models yielded successful fitting,
we adopted the one with the smaller $\chi^{2}$ value. The result is given in
figure~\ref{f6} for the spectra and best-fit models and in table~\ref{t06} for the
best-fit parameters.

\input{t07}

The composite spectra show signatures of the Fe K line feature. We derived their
equivalent width (EW) by locally fitting the spectra in the 4--8~keV range with a
power-law and a Gaussian line. With the energy resolution of Chandra, we cannot separate
the 6.4, 6.7, and 7.0 keV lines, thus we treated them as a single emission line. The
free parameters are the photon index and flux of the power-law and the line center and
the normalization of the Gaussian component. The line width was fixed to 0.01 keV. We
tested the necessity of the Gaussian component using the F-test and found that it was
indeed required for Group A and B spectra with significances of 87.5\% and 91.6\%
retrospectively. On the other hand, the significance of F-test of Group C is low,  
indicating that it does not require a significant Gaussian component. This can be attributed to the lack of 
the photon statistics in Group C. The result is shown in table~\ref{t07}.

We further obtained the X-ray surface brightness  2.7$\times$10$^{-11}$ $\ergcmsdeg$ in 2--8 keV by averaging
all the X-ray point sources identified in NIR.  From \citet{Iso2012}, the entire X-ray
surface brightness of the GBXE in the same band is 1.01$\times$10$^{-10}$ $\ergcmsdeg$. 
As a result, we estimate that X-ray surface brightness from the X-ray point sources with NIR counterpart candidates in this 
and our previous studies account for $\sim$26\% of the entire X-ray surface brightness of the GBXE.

%%%%%%%%%%%%%%%%%%%%%%%%%%%%%%%%%%%%%%%%%%%%%%%%%%%%%%%%%%%%%%%%%%%%%%%%%%%%%%%%%%
\subsubsection{X-ray variability}\label{s4-1-3}
%%%%%%%%%%%%%%%%%%%%%%%%%%%%%%%%%%%%%%%%%%%%%%%%%%%%%%%%%%%%%%%%%%%%%%%%%%%%%%%%%%%
We investigated temporal behavior of the X-ray sources with NIR counterparts. For all
the X-ray sources with NIR counterparts, we applied the Kolmogorov-Smirnov (KS) test
against the null hypothesis that the X-ray light curves are constant. We consider that
the sources with the null hypothesis probability ($P_{\mathrm{K-S}}$) of $< 5 \times
10^{-3}$ as ``definitely variable" and those with $5 \times 10^{-3} < P_{\mathrm{K-S}} <
5 \times 10^{-2}$ as ``possibly variable". As a result, 11 sources are definitely
variable and 19 are possibly variable (table~\ref{t08}). Almost all the variable sources
belong to Group B or C, and more than a half belong to Group C.

\input{t08}

\bigskip

%%%%%%%%%%%%%%%%%%%%%%%%%%%%%%%%%%%%%%%%%%%%%%%%
\subsubsection{X-ray to NIR flux ratio}\label{s4-1-4}
\begin{figure*}[htbp]
\begin{center}
 \includegraphics[width=1.0\textwidth, clip]{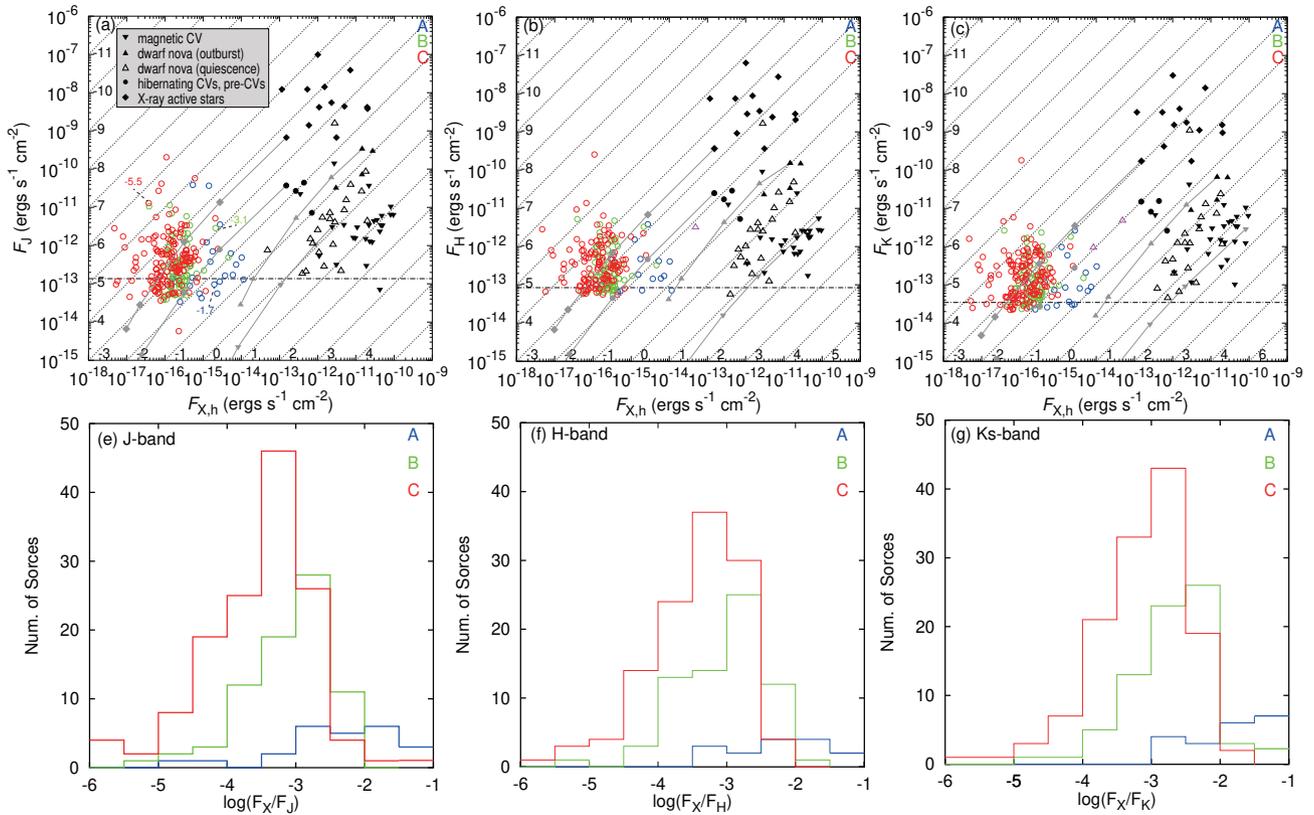}
 \end{center}
 \hspace{50cm}
 \scriptsize
 \vspace{-4mm}
 \caption{ (Top) Flux ratio between the hard (2--8~keV) X-ray and (a) \ji-band, (b)
 \hi-band, or (c) \ksi-band flux of the X-ray sources with NIR counterparts. The
 dotted lines indicate the iso-flux-ratio with its logarithmic values given along
   the $x$ and $y$ axes. Different colors are
 assigned based on the grouping in \S~\ref{s3-3}. The purple triangle shows hard X-ray
 sources with NIR emission features (\S~\ref{s4-1-6}).
 The horizontal dashed-line in the panels of the upper figures indicate the 90\% completeness limit of each band.
 Representative sources of the following categories are shown for comparison in different symbols in black
 as in the legend in (a); (i) magnetic CVs (polars ; \cite{Ezuka1999,Mukai2017,
 Xu2016} and intermediate polars : \cite{Ezuka1999, Mukai2017, Xu2016}), 
 (ii) dwarf novae (\cite{Baskill2005,Wada2017,Ebisawa2005, Xu2016, Wada2017}), 
 (iii) hibernating CVs, pre-CVs (\cite{Matranga2012, Schwope2014}), 
 (iv) active stars (during flares ; \cite{Pandey2008,Pandey2012, Tsuboi2016} and in quiescence ;
 (\cite{Pandey2008}),
 For several sources, we also showed the expected trajectories when the distance (and
 proportionally the extinction column proportional to the distance) increases as
 \nh = 1$\times$10$^{21}$ ($\sim$350 pc), 3$\times$10$^{21}$ ($\sim$1 kpc), 1$\times$10$^{22}$ ($\sim$3 kpc),
 1.5$\times$10$^{22}$ ($\sim$5 kpc) with the solid lines from upper-right to lower-left.
The relation between the column density and the distance is based on the
 typical hydrogen column density $\sim$1 cm$^{-3}$ of the interstellar medium.
 We used the \texttt{wabs} model (\cite{Morrison1983}) with the abundance table (\cite{Anders1989}) for the extinction calculation.
 (Bottom) distribution of the \fx/\fnir~values in the (d) \ji- (e) \hi-, and (f) \ksi-bands.
 }
 \label{f7}
\end{figure*}

%%%%%%%%%%%%%%%%%%%%%%%%%%%%%%%%%%%%%%%%%%%%%%%%
We investigated ratio of the X-ray flux to the NIR flux. The ratio is expected to be relatively high
for the binaries containing compact objects and low for stellar sources. 
We employed the photometric X-ray flux at 2--8 keV (table~\ref{t03} column 9) estimated from the count rates and the
median energies \citep{Morihana2013}. Figure~\ref{f7} top panel shows the ratio between
the hard (2--8~keV) X-ray ($F_{\mathrm{X,h}}$) against NIR flux ($F_{\mathrm{NIR}}$) for
all the X-ray sources with NIR counterparts. 
$F_{\mathrm{NIR}}$ is defined as $F_{\mathrm{0\lambda}}d\lambda10^{-0.4m}$,
where $F_{\mathrm{0\lambda}}$, $d\lambda$, and $m$ are the flux at 0 mag for each band, the band width, and 
the Vega magnitude, respectively.
The logarithm of the ratio
($\log{F_{\mathrm{X,h}}/F_{\mathrm{NIR}}}$) shows different distributions among the three groups, 
which are more influenced by the differences of $F_{\mathrm{X,h}}$ than that of $F_{\mathrm{NIR}}$.
This is depicted in figure~\ref{f7} bottom panels, which show the histograms of
$F_{\mathrm{X,h}}/F_{\mathrm{NIR}}$ values. For example, in the
$\log{F_{\mathrm{X,h}}/F_{Ks}}$ plot (figure~\ref{f7}c), sources span in the range of
--3 to --1 (Group A), --4 to --2 (Group B), and from --6 to --2 (Group C).  We also
plotted the flux ratios of several representative sources with known nature that are
for candidates of the X-ray point sources constituting the GBXE.

%\begin{landscape}
%\end{landscape}

%%%%%%%%%%%%%%%%%%%%%%%%%%%%%%%%%%%%%%%%%%%%%%%%%%%%%%%%%%%%%%%%%%%%%%%%%%%%%%%%
\subsubsection{Distance and X-ray luminosity}\label{s4-1-5}
\begin{figure}[htbp]
 \begin{center}
  \includegraphics[width=0.45\textwidth]{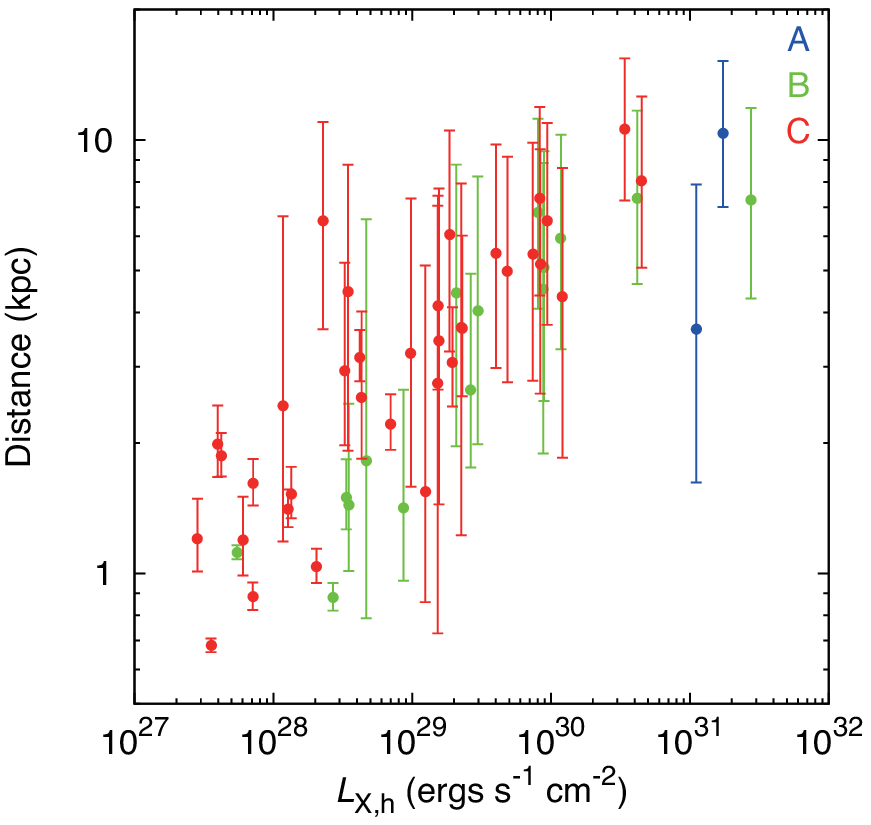}
 \end{center}
 \vspace{-0.1cm}
 \caption{Scatter plot of the distance by Gaia (\cite{Bailer2018}) and the 2--8 keV X-ray luminosity
  for the sources with distance estimates. }
 \label{f8}
\end{figure}

%%%%%%%%%%%%%%%%%%%%%%%%%%%%%%%%%%%%%%%%%%%%%%%%%%%%%%%%%%%%%%%%%%%%%%%%%%%%%%%%%%%%
Almost all the previous works discussing X-ray point sources constituting the GBXE and GRXE
were based on the flux without knowing the distance. Now, we have identified 109 Gaia
counterparts ($\S~\ref{s4-1-1}$). Among them, 61 sources have parallax measurements
\citep{Bailer2018}. The breakdown is that 2, 15, and 44 sources belong to the Group A,
B, and C, respectively (table~\ref{t05}). We derived their distance and X-ray luminosity
(\lx) in the 2--8 keV band 
without correcting the interstellar absorption. The distance and luminosity
distributions are shown in figure~\ref{f8}. 
We found that  most (if not all, given the large distance errors) X-ray sources with Gaia counterparts are within  4 kpc 
 thus are foreground objects to the GBXE. Here, one should keep in mind that the false positive rates among the Gaia matches are
 from 7\% (Group A) to 30\% (Group C).
 The average X-ray luminosity of each group is $\sim 10^{30.4^{+1.2}_{-0.5}} \ergs$ (Group A), 
$\sim 10^{29.6^{+0.3}_{-0.8}} \ergs$ (Group B), and $\sim$ 10$^{29.2^{+0.4}_{-0.4}}$ $\ergs$ (Group C).

%%%%%%%%%%%%%%%%%%%%%%%%%%%%%%%%%%%%%%%%%%%%%%%%%%%%%%%%%%%%%%%%%%%%%%%%%%%%%%%%
\subsubsection{NIR spectroscopy results}\label{s4-1-6}
\begin{figure*}[htbp]
 \begin{center}
  \includegraphics[width=0.85\textwidth]{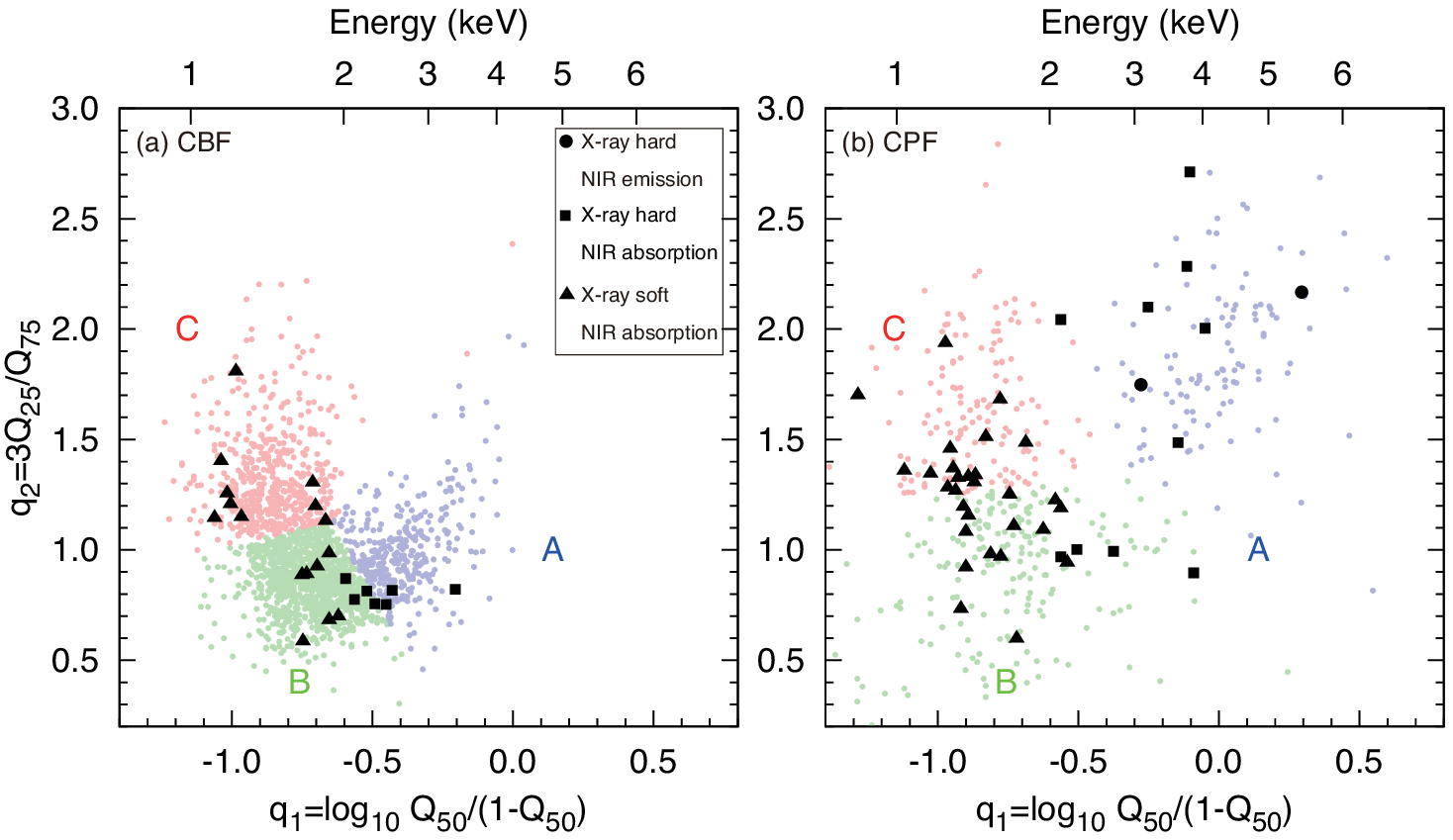}
 \end{center}
 \vspace{-0.1cm}
 \caption{Quantile distribution of all the X-ray sources (pale colors) and those with NIR
 spectroscopic results in the CBF (left) and the CPF
 (right). Black filled circles, squares, and triangles respectively indicate X-ray hard
 sources with NIR emission features (H$+$emi in table~\ref{t09}), X-ray hard sources
 with NIR absorption features (H$+$abs), and X-ray soft sources with NIR absorption
 sources (S$+$abs). The typical error of q1 and q2 are 0.02 and 0.06, respectively, 
 for a source with average 40 counts. The minimum and 
 maximum numbers of the counts for the
 X-ray sources with NIR spectra are 12 counts and 222 counts, respectively.}\label{f9}
\end{figure*}

%%%%%%%%%%%%%%%%%%%%%%%%%%%%%%%%%%%%%%%%%%%%%%%%%%%%%%%%%%%%%%%%%%%%%%%%%%%%%%%%%%%%
In our previous work \citep{Morihana2016}, we carried out NIR spectroscopy for 23 of the
present sources in the CBF and 40 sources in the CPF. To compensate for the paucity of the statistics
in the CBF alone, we also refer to the result in the CPF here. To facilitate the
comparison between the two fields, we constructed the $q_1$-$q_2$ diagram for the CPF (figure~\ref{f9}) and
classified the NIR-identified X-ray sources into Groups A--C using the same algorithm
described in \S~\ref{s3-3}. Due to the difference of the extinction toward CBF and CPF,
the scatter of the data is different, but the overall trend is similar to each other.

Three distinctive types were found based on the combination of X-ray spectral hardness
and NIR line features; hard X-ray sources with NIR emission features of
H\emissiontype{I} and He\emissiontype{II} (H+emi), hard X-ray sources with NIR
absorption features (H+abs), and soft X-ray sources with NIR absorption features
(S+abs).  Here, the X-ray spectral hardness is defined by hardness ratio as
$(H-S)/(H+S)$ where $H$ and $S$ are the normalized count rates in the hard (2--8 keV)
band and soft (0.5--2 keV) band.  We show their distribution in the $q_1$-$q_2$ diagram
in figure~\ref{f9} and statistics in table~\ref{t09}. It should be noted that the NIR features, which are
unrelated to $q_1$ or $q_2$ by definition, are distributed in a biased manner in the
diagram. For example, the hard X-ray sources with NIR emission features (H$+$emi, only
found in CPF) belong to Group A. Group C sources only consist of the soft X-ray
sources with NIR absorption features (S$+$abs).

\input{t09}

%%%%%%%%%%%%%%%%%%%%%%%%%%%%%%%%%%%%%%%%%%%%%%%%%%%%%%%%%%%%%%%%%
\subsection{Major populations}\label{s4-2}
%%%%%%%%%%%%%%%%%%%%%%%%%%%%%%%%%%%%%%%%%%%%%%%%%%%%%%%%%%%%%%%%%%
Based on the characterizations in various aspects shown in \S~\ref{s4-1}, we summarize
distinctive characteristics of the sources
in simplicity order of the features--- Group C, A, and B.

%%%%%%%%%%%%%%%%%%%%%%%%%%%%%%%%%%%%%%%
\subsubsection{Group C}\label{s4-2-1}
%%%%%%%%%%%%%%%%%%%%%%%%%%%%%%%%%%%%%%%
The first is the Group C sources, which have the following properties
distinctive from the others:

\begin{enumerate}
 \renewcommand{\labelenumi}{(\arabic{enumi})}
 \item The composite X-ray spectrum is represented by a two-temperature thermal
       plasma model with temperatures of \kt$\sim$0.6 and 2.0 keV
       (table~\ref{t06}). Unlike the other groups, Fe K line was not observed
       significantly (table~\ref{t07}).
 \item The fraction of time-variable sources is the largest (table~\ref{t08}).
 \item The ratio of X-ray to NIR flux is the lowest in the range of 10$^{-6}$ to
       10$^{-2}$ (figure~\ref{f7}).
 \item The fraction of optically-identified sources is the largest
       (table~\ref{t05}). Most sources are located within 4 kpc
       (figure~\ref{f8}). The average X-ray luminosity is the lowest,
       1.57$\times$10$^{29}$$\ergs$ in the 2--8 keV band.
 \item All the Group C sources studied in the NIR spectroscopy exhibit
       absorption features typical of late-type stars \citep{Morihana2016}.
\end{enumerate}

These characteristics are most consistent as the major population of the Group C sources
being the X-ray active stars. The plasma temperatures and X-ray luminosity are consistent with
the values typically seen in X-ray active stars such as RS CVn systems
\citep{Dempsey1993,Gudel1999,Covino2000,Pandey2012}.  The ratio of X-ray to NIR flux 
that mainly driven by the differences in  $F_{\mathrm{X,h}}$
is also consistent with those of representative sources (figure~\ref{f7}). The high fraction of the X-ray
variable sources is understood, because
the X-ray active stars show occasional flare-like
activities.  Average X-ray luminosity of the optically-identified sources is consistent
with that of the typical value (10$^{27}$--10$^{30}$ $\ergs$; \cite{Schmitt1990}) of
X-ray active stars. From the above characteristics, we argue that 
the major population of the Group C sources is X-ray active stars.

%%%%%%%%%%%%%%%%%%%%%%%%%%%%%%%%%%%%%%%
\subsubsection{Group A}\label{s4-2-2}
%%%%%%%%%%%%%%%%%%%%%%%%%%%%%%%%%%%%%%%
We next discuss the Group A sources, which  have the following characteristics:

\begin{enumerate}
 \renewcommand{\labelenumi}{(\arabic{enumi})}
 \item The composite X-ray spectrum is described by the combination of a thermal plasma
       model with \kt$\sim$4.6 keV and a power-law model with $\Gamma=$0.25
       (table~\ref{t06}). The Fe K line was detected with the center at 6.7~keV and the
       EW of 162~eV (table~\ref{t07}).
 \item None but one source shows X-ray variability (table~\ref{t08}).
 \item The ratio of X-ray to NIR flux is the largest with 10$^{-3}$ to 10$^{-1}$
       (figure~\ref{f7}).
 \item The fraction of optically-identified sources is the smallest
       (table~\ref{t05}). The average X-ray luminosity of the optically-identified
       sources is $\sim$10$^{30}$~$\ergs$ in the 2--8~keV band, the largest among the three groups.
 \item All the hard X-ray sources with NIR emission features belong to
       Group A. Group A sources also include hard X-ray sources with NIR absorption
       features (table~\ref{t09}).
\end{enumerate}

We interpret these features that the Group A sources are a mixture of the magnetic and
non-magnetic CVs. The ratio of the X-ray to NIR flux (figure~\ref{f7}) 
%\textcolor{red}{that mainly driven by the differences in  $F_{\mathrm{X,h}}$}
is consistent with those of representative magnetic CVs and dwarf novae. They are time
variable from their names, but the occurrence of such variability is less frequent than
flares in X-ray active stars and its duration is long, hence a fraction of the X-ray
variable sources is considered small. All the sources with the NIR emission line features of
H\emissiontype{I} and He\emissiontype{II} belong to this group, which is consistent with
the idea that the Group A sources are those with prominent accretion disks.
Those without the emission features are presumably sources with an accretion rate 
that is too small to exhibit prominent emission lines over the stellar photospheric absorption features.

The typical EW of the Fe K lines (6.4, 6.7, and 7.0~keV combined) is $\sim$350~eV for
magnetic CVs and $\sim$700~eV for non-magnetic CVs \citep{Koyama2018}. The EW of the
composite spectrum of the Group A sources is smaller. Therefore,  contribution of the
non-magnetic CVs is considered sub-dominant.

%%%%%%%%%%%%%%%%%%%%%%%%%%%%%%%%%%%%%%
\subsubsection{Group B}\label{s4-2-3}
%%%%%%%%%%%%%%%%%%%%%%%%%%%%%%%%%%%%%
Finally, we discuss Group B sources with the following characteristics:
\begin{enumerate}
 \renewcommand{\labelenumi}{(\arabic{enumi})}
 \item Some sources exhibit X-ray variability (table~\ref{t08}).
 \item Ratio of theX-ray to NIR flux is in the range of 10$^{-4}$ to 10$^{-2}$, between Group A and C
       (figure~\ref{f7}).
 \item Fraction of the optically-identified sources is the second largest next to Group
       C (table~\ref{t05}). The average X-ray luminosity is 3.7$\times$10$^{29}$~$\ergs$
       in the 2--8~keV band, the second largest next to Group A.
 \item All the sources with the NIR spectra exhibited absorption features typical of the late-type 
       stars \citep{Morihana2016}.
 \item The composite X-ray spectrum is described by a two-temperature thermal plasma
       model with temperatures of \kt$\sim$0.7 and 6.3 keV (table~\ref{t06}). The Fe K
       line is the strongest with an EW of $\sim$500 eV (table~\ref{t07}).
\end{enumerate}

Most of these characteristics are in between Group A and C as expected from the
definition in figure~\ref{f5}. However, we do not consider that the Group B sources are a
mere mixture of the major populations of the Group A and C sources. This is because the
Fe K line EW of the Group B average X-ray spectrum is the largest of the three groups,
which cannot be explained by the combination of Group A and C. We speculate that,
unlike Group A, non-magnetic CVs are dominant in Group B.

However, the Group B sources also have the characteristics that cannot be explained
by \textit{known} non-magnetic CVs (or, sources identified as dwarf novae) alone. The
ratios of the X-ray to NIR fluxes 
and the typical X-ray luminosities are both smaller than
those of dwarf novae (figure~\ref{f7}). NIR spectra of dwarf novae often exhibit
H\emissiontype{I} emission lines \citep{Dhillon2000} from their accretion disks with a
typical accretion rate of $\dot{M} \sim 10^{-10}-10^{-11}M_\odot$~yr$^{-1}$, but \textit{no
sources in Group B show NIR emission lines.}

This can be reconciled if we assume WD binaries that are not yet recognized as
dwarf novae due to absence of the dwarf nova activities. We could also consider
other types of WD binaries such as detached WD binaries \citep{Shara1986, Warner2003},
hibernating CVs, and precursors of CVs called ``pre-CVs'' \citep{Warner1995}. We hereafter call them all together
as ``quiet'' WD binaries.  The ''quiet'' WD binaries have  low accretion mass rates less
than $10^{-11}M_\odot$~yr$^{-1}$.  Without dwarf novae activities, these sources are
almost indistinguishable from the field main sequence stars in optical and NIR photometric
observations. In X-rays, however, some of these sources do show features as WD
binaries with \lx$\sim$10$^{29}$~$\ergs$ (2--10 keV) and the plasma temperature being
$\sim$4 keV \citep{Matranga2012,Schwope2014}, which are typical of Group B sources
though the average X-ray luminosity of Group B has a large error. 
We thus speculate that Group B sources consist mainly of the quiet WD binaries.

\bigskip

\begin{figure*}[htbp]
 \begin{center}
  \includegraphics[width=0.8\textwidth]{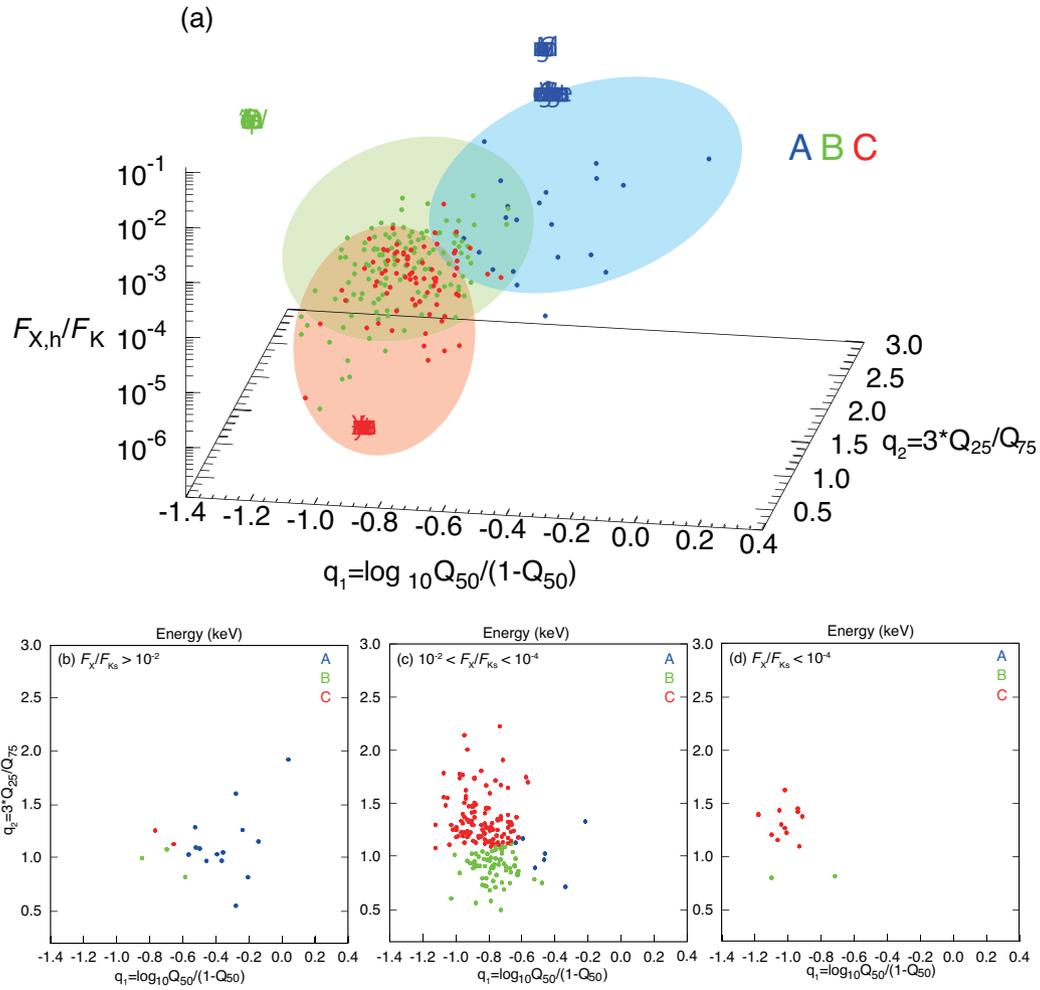}
 \end{center}
 \vspace{-0.1cm}
 \caption{(a) Three-dimensional plot of $q_1$, $q_2$, and the ratio of X-ray to NIR
 flux. Their cross-section on the $q_1$--$q_2$ plane for the ratio of X-ray to NIR flux of (b) $>$ 10$^{-2}$, 
 (c) 10$^{-4}$--10$^{-2}$, and (d) $<$ 10$^{-4}$.}
 \label{f10}
\end{figure*}

To summarize, we suggest that the dominant population of each group as follows:
X-ray active stars (Group C), magnetic-CVs and non-magnetic CVs with high accretion rates 
(Group A), and ``quiet'' WD binaries (Group B). These populations in the groups
overlap with each other in figure~\ref{f9}, in which only the X-ray properties are
used. The overlap can be disentangled to some extent if we add another axis using NIR
property. Figure~\ref{f10} shows the three-dimensional plot by adding the X-ray to NIR
flux ratio. Group A sources tend to have small values along the X-ray to NIR flux ratio, while 
Group C sources have large values.

%%%%%%%%%%%%%%%%%%%%%%%%%%%%%%%%%%%%%%%%%%%%%%%%%%%%%%%%%%%%%%%%%%%%%%%%
\subsection{Populations contributing to the GBXE}\label{s4-3}
%%%%%%%%%%%%%%%%%%%%%%%%%%%%%%%%%%%%%%%%%%%%%%%%%%%%%%%%%%%%%%%%%%%%%%%%
In previous studies, magnetic CVs and X-ray active stars have been suggested as two most popular groups
to explain the GBXE and GRXE. For example, \citet{Revnivtsev2006} discussed the
X-ray population based on X-ray luminosity function in the range of \lx$\sim
10^{30}-10^{32} \ergs$ (2--10 keV) and concluded that the major population constituting
the GRXE is X-ray active binaries,  which dominate the population below
\lx$\sim$10$^{31}$$\ergs$. \citet{Yuasa2012} constructed a wide-band X-ray spectral model
of a type of magnetic CVs (intermediate polars) and fitted it to the 2--50 keV spectra of the
GRXE obtained with Suzaku. They argued that the GRXE above 10 keV can be mostly
explained by magnetic CVs and that below 10 keV mostly by X-ray active stars. \citet{Hong2012}
concluded that the hard X-ray flux above 3 keV is dominant by magnetic CVs and 
contribution of the X-ray active stars is less than 20\%,  based on reanalysis of the X-ray
data in the CBF.

Most recent studies propose some revisions on these pictures. Though magnetic CVs and X-ray
active stars are major populations, there should be other classes of sources
that are more dominant. Following the work by \citet{Mukai1993} based on the X-ray
luminosity function, \citet{Nobukawa2016}, \citet{Yamauchi2016}, and \citet{Xu2016}
suggested that non-magnetic CVs should be a more dominant population than magnetic CVs and
X-ray active stars. They have reached this conclusion by comparing the EW, the line
ratio, and the scale height of the Fe K line complex between the GRXE and GBXE and the
composite X-ray spectra of dwarf novae.

Our results are in line with the conclusions in these recent works in 
that the dominant contributor to the GRXE and GBXE are non-magnetic CVs.
What we claim in this paper is that \textit{the major
contributor to the GBXE is quiet WD binaries that are yet to be recognized
as dwarf novae}. In fact, such sources are predicted to outnumber the known population of
dwarf novae. Theories of CV evolution (e.g., \cite{Howell2001,Hellier2001,Warner1995}) indicate
that CVs remain in the state of low accretion rates for most of their lifetime. An
increasing number of such sources are now identified spectroscopically as byproducts of
the Sloan Digital Sky Survey and their follow-up studies (Szcody et
al.~\yearcite{Szkody2002,Szkody2003,
Szkody2004,Szkody2005,Szkody2006,Szkody2007,Szkody2009,Szkody2011}). X-ray emission is
detected from some of these sources \citep{Reis2013} with an average X-ray luminosity of as low as
$8 \times 10^{29}$~$\ergs$ in the 0.5--10~keV. The luminosity is much lower than that
of typical dwarf novae \citep{Byckling2010}. \citet{Reis2013} further argue that these
faint sources are more representative of the WD binaries than those in the brightest end
(mostly known as dwarf novae) and that the faint sources account for a significant part of the GRXE. Our
findings agree with their claims.

%%%%%%%%%%%%%%%%%%%%%%%%%%%%%%%%%%
\section{Summary}\label{s5}
%%%%%%%%%%%%%%%%%%%%%%%%%%%%%%%%%%%
Main results of the present paper are as follows.

\begin{enumerate}
 \item We carried out a deep NIR \ji\hi\ksi-band imaging observation at the center of the
       deep CBF using MOIRCS on the 8.2~m Subaru telescope,  and identified $\sim$50\% 
       of the X-ray point sources with NIR counterpart candidates. 
       The false positive rate among the identifications is estimated to be 25\%.
       The identification rate increased significantly from our previous NIR survey by using the 1.4~m IRSF telescope in
       the same field (11\%) \citep{Morihana2012}.
 \item We classified the X-ray sources into three (A, B, and C) groups based on their
       X-ray colors. We generated three-dimensional plots (X-ray color-color and 
       the ratio of the X-ray to the NIR flux) to disentangle the overlap among the groups.
       We have characterized the overall features of each group based on
       their NIR identification rates, X-ray composite spectra, the flux ratios between the X-ray
       and NIR bands, X-ray time variability, X-ray average luminosities, and NIR
       spectroscopy results.
 \item Based on the overall characteristics, we argued that the dominant populations of
       each group are as follows: magnetic-CVs and non-magnetic CVs with high accretion
       rates (Group A), ``quiet'' WD binaries of low mass accretion rates (Group B), and X-ray 
       active stars (Group C).
 \item The composite X-ray spectra of Group B sources have the largest Fe K equivalent
       width comparable with that of the GBXE. This leads us to speculate that the major constituent
       of the faint discrete sources for the GBXE is ``quiet'' WD binaries with low mass accretion rates.
\end{enumerate}

%%%%%%%%%%%%%%%%%%%%%%%%%%%%%%%%%%%%%%%%%%%%%%%%%%%%%%%%%%%%
% Acknowledgments
%%%%%%%%%%%%%%%%%%%%%%%%%%%%%%%%%%%%%%%%%%%%%%%%%%%%%%%%%%%%
\medskip
\begin{ack}
 We thank Ichi Tanaka for providing the MOIRCS data reduction pipeline as well as
 useful discussion. We acknowledge the reviewer for useful comments. This work is based on the data collected at Subaru Telescope operated by
 the National Astronomical Observatory of Japan, public data obtained from Chandra X-ray
 Center which is operated for NASA by the Smithsonian Astrophysical Observatory, the VVV
 Survey which is supported by the European Southern Observatory, and the data of the
 European Space Agency (ESA) space mission Gaia. Gaia data are being processed by the
 Gaia Data Processing and Analysis Consortium (DPAC). Funding for the DPAC is provided
 by national institutions, in particular the institutions participating in the Gaia
 Multi-Lateral Agreement (MLA).
 IRAF is distributed by the National Optical Astronomy Observatories, which are operated
 by the Association of Universities for Research in Astronomy, Inc., under cooperative
 agreement with the National Science Foundation.
 This research has made use of SAOImage DS9, developed by Smithsonian Astrophysical
 Observatory. This work has also made use of software from High Energy Astrophysics
 Science Archive Research Center (HEASAC) which is provided by NASA Goddard Space Flight
 Center. 
 K.\,M. is financially supported by MEXT/JSPS KAKENHI Grant Numbers 17K18019 and 19H01939.
 \end{ack}

\bibliographystyle{aa}
\bibliography{ms}

\end{document}

%% file: t01.tex
\setlength{\tabcolsep}{0.1in}
\begin{table*}[htbp]
\small
 \caption{Subaru observation log.}\label{t01}
%\begin{center}
  \begin{tabular}{llllll}
   \hline
   \hline
    Object & R.\,A. (J2000)         & Dec. (J2000)                     & Exp. time (min)  & Obs. date & Airmass \\
   \hline
     Chandra Bulge Field (CBF)   & 17:51:27.86   & $-$29:35:31.40  &  25 (\ji, \ksi), 30 (\hi)       & 2012-05-08     &  1.535--1.762\\
     QUARET (\ji-band sky)         & 18:36:10.28   & $-$07:07:10.29        &         49                                    & 2012-06-11     &   1.236--1.551 \\
     G24.47+0.49 (\hi-band sky)  & 18:34:06.34   & $-$07:19:39.82        &        40      & 2012-06-11     &    1.125--1.134  \\
     SSA22 (\ksi-band sky)    & 22:17:22.58   &  $+$00:17:55.15    &         8             & 2012-06-11     &    1.061               \\
     \hline
  \end{tabular}
%\end{center}
%\par \noindent
%\footnotemark[$*$]
\end{table*}

%% file: t02.tex
\setlength{\tabcolsep}{0.015in}
\begin{table}[htbp]
\small
 \caption{Summary of NIR identification.}\label{t02}
%\begin{center}
  \begin{tabular}{lllllll}
   \hline
   \hline
   \multicolumn{1}{c}{}&
   \multicolumn{3}{c}{Num. of sources} &
        \multicolumn{3}{c}{NIR ID Rate} \\
\hline
        & \ji & \hi & \ksi & \ji & \hi & \ksi\\
   \hline
     X-ray Point Sources\footnotemark[$*$]        & 533 & 539 & 530  & \phantom{0}-- & \phantom{0}-- & \phantom{0}--\\
     2MASS \& SIRIUS IDed\footnotemark[$\dagger$] & \phantom{0}49  & \phantom{0}46 & \phantom{0}44 & \phantom{0}9\% & \phantom{0}9\% & \phantom{0}8\%\\
     MOIRCS IDed                              & 206 & 194 & 205     & 39\% & 36\% & 39\%\\
     2MASS+SIRIUS+MOIRCS IDed                 & 255 & 240 & 249     & 48\% &  45\% & 47\%\\
     \hline
  \end{tabular}
%\end{center}
\par \noindent
\footnotemark[$*$]Number of the X-ray point sources detected by \citet{Morihana2013} within the image of
 each band.
\footnotemark[$\dagger$]Number of the X-ray point sources with candidate 2MASS or SIRIUS counterparts (\cite{Morihana2012}).
\end{table}

%% file: t03.tex
\setlength{\tabcolsep}{0.035in}
\begin{table*}
\caption{NIR identified X-ray source catalog.\label{t03}}
\scriptsize
\begin{minipage}{\textwidth}
\begin{center}
\begin{tabular}[14]{ccccccccccccccc}
%\begin{center}
  \hline
  \hline
        \multicolumn{2}{c}{Source} &
        \multicolumn{5}{c}{Position} &
        \multicolumn{2}{c}{Characteristics} &
        \multicolumn{6}{c}{Photometry}\\
        \multicolumn{2}{c}{\hrulefill} &
        \multicolumn{5}{c}{\hrulefill} &
        \multicolumn{2}{c}{\hrulefill} &
        \multicolumn{6}{c}{\hrulefill} \\
  Number  &  CXOU J & 
  $\rm{R.\,A.(X)}$ & $\rm{Dec.(X)}$ & $\rm{R.\,A. (NIR)}$ & $\rm{Dec. (NIR)}$ & Err
 & Phot\fx~(0.5--8 keV) & Phot\fx~(2--8 keV) & 
  \ji & \ji err & \hi & \hi err & \ksi & \ksi err\\
  \#  &  & 
  (deg) & (deg) &  (deg) & (deg) &(\arcsec)
  & (ergs cm$^{-2}$ s$^{-1}$) &  (ergs cm$^{-2}$ s$^{-1}$) & 
  (mag)& (mag)& (mag)& (mag)& (mag)& (mag)\\
\hline
    (1) & (2) 
  & (3) & (4) &(5) & (6) & (7)
  & (8) & (9) 
  & (10) & (11)& (12)& (13) & (14) & (15)\\
\hline
1000 & 175128.13$-$293703.7  &267.867210 &-29.61772 &267.86730 &-29.61760 & 0.1 & 2.9$\times$10$^{-16}$ & 6.8$\times$10$^{-17}$ & 11.26 & 0.01 & 10.95 & 0.01 & 10.92 & 0.01\\ 
1004 & 175128.20$-$293347.0  &267.867530 &-29.56306 &267.86750 &-29.56310 & 0.1 & 8.0$\times$10$^{-16}$ & 5.4$\times$10$^{-16}$ & 10.86 & 0.01 & 0.00 & 0.00 & 0.00 & 0.00\\ 
1006 & 175128.25$-$293248.3  &267.867730 &-29.54677 &267.86768 &-29.54680 & 0.2 & 1.5$\times$10$^{-16}$ & 1.3$\times$10$^{-16}$ & 17.97 & 0.12 & 0.00 & 0.00 & 16.63 & 0.10\\ 
1010 & 175128.35$-$293621.5  &267.868140 &-29.60599 &267.86810 &-29.60610 & 0.1 & 1.8$\times$10$^{-16}$ & 1.2$\times$10$^{-16}$ & 15.12 & 0.10 & 14.20 & 0.12 & 14.20 & 0.14\\ 
1011 & 175128.36$-$293714.7  &267.868170 &-29.62076 &267.86811 &-29.62069 & 0.1 & 1.5$\times$10$^{-16}$ & 1.2$\times$10$^{-16}$ & 14.96 & 0.01 & 14.30 & 0.01 & 14.13 & 0.01\\ 
1013 & 175128.38$-$293438.2  &267.868290 &-29.57729 &267.86826 &-29.57723 & 0.1 & 1.2$\times$10$^{-16}$ & 6.4$\times$10$^{-17}$ & 16.65 & 0.04 & 16.04 & 0.05 & 15.82 & 0.04\\ 
1014 & 175128.39$-$293215.4  &267.868330 &-29.53762 &267.86837 &-29.53753 & 0.2 & 2.5$\times$10$^{-16}$ & 2.4$\times$10$^{-16}$ & 17.53 & 0.09 & 16.63 & 0.08 & 15.97 & 0.07\\ 
1019 & 175128.48$-$293625.0  &267.868680 &-29.60696 &267.86860 &-29.60690 & 0.1 & 1.4$\times$10$^{-16}$ & 4.7$\times$10$^{-18}$ & 13.71 & 0.03 & 12.66 & 0.02 & 12.32 & 0.02\\ 
1020 & 175128.52$-$293328.7  &267.868870 &-29.55798 &267.86888 &-29.55798 & 0.0 & 1.0$\times$10$^{-15}$ & 4.7$\times$10$^{-16}$ & 15.61 & 0.03 & 14.99 & 0.03 & 14.71 & 0.02\\ 
1022 & 175128.57$-$293220.9  &267.869060 &-29.53916 &267.86892 &-29.53925 & 0.2 & 1.6$\times$10$^{-16}$ & 1.1$\times$10$^{-16}$ & 15.34 & 0.01 & 0.00 & 0.00 & 14.71 & 0.02\\ 
\hline
\end{tabular}
\end{center}
\scriptsize
\vspace{-1mm}
\footnotetext{
Col. (1): X-ray catalog source number in \citet{Morihana2013}, sorted by R.\,A.
Col. (2): IAU designation. 
Col. (3)--(6) : R.\,A. and Dec.\ in the equinox J2000.0 by X-ray and NIR.
Col. (7): X-ray position error (1 $\sigma$).
Col. (8), (9) : Photometric flux in the total (0.5--8.0~keV) and the hard (2.0--8.0 keV) band. The photometric flux is defined as count rate 
     multiplied by the median energy divided by the average effective area (\cite{Tsujimoto2005, Morihana2013}).
Col. (10)--(15): NIR magnitude and error. (A portion of the catalog is shown here for guidance regarding its form and content.
The full version is shown in a machine-readable form in the online journal.)
}
\end{minipage}
\end{table*}

%% file: t04.tex
\setlength{\tabcolsep}{0.14in}
\begin{table*}[htbp]
 \caption{Estimates of false positive detections.}\label{t04}
 %\begin{center}
 \small
 \begin{tabular}{cccccccccc}
  \hline
  \hline
  Range  & \multicolumn{3}{c}{$N_{\mathrm{X,MOIRCS}}$\footnotemark[$*$]} & \multicolumn{3}{c}{$\Sigma$\footnotemark[$\dagger$] (arcsec$^{-2}$)} & \multicolumn{3}{c}{$N_{\mathrm{FP}}$\footnotemark[$\ddagger$]} \\
  (mag)  & \ji & \hi & \ksi & \ji & \hi & \ksi & \ji & \hi & \ksi \\
  \hline
  14--15 & 29 & 48 & 59 & 0.022 & 0.024 & 0.026 & \phantom{0}9.2 (32\%) & 10.2 (21\%) & 10.8 (18\%)\\
  15--16 & 61 & 56 & 58 & 0.023 & 0.031 & 0.034 & \phantom{0}9.6 (16\%) & 13.1 (23\%) & 14.1 (24\%)\\
  16--17 & 58 & 63 & 62 & 0.038 & 0.048 & 0.040 & 15.9 (27\%)& 20.3 (32\%) & 16.6 (27\%)\\
  17--18 & 60 & 21 & 12 & 0.049 & 0.033 & 0.019 & 20.5 (34\%)& 14.0 (67\%)& \phantom{0}7.9(66\%)\\
  \hline
 \end{tabular}
\par \noindent
 \footnotemark[$*$] Number of MOICRS-identified X-ray sources.
 \footnotemark[$\dagger$] Surface number density of MOIRCS sources.
 \footnotemark[$\ddagger$] Estimated number of false positive detections and false positive rate.
 The number in parentheses represents the false positive rate.
\end{table*}

%% file: t05.tex
\setlength{\tabcolsep}{0.028in}
\begin{table}[htbp]
\small
 \caption{Summary of identification in each group.}\label{t05}
%\begin{center}
  \begin{tabular}{ccccc}
   \hline
   \hline
    Group       & Num. of X-ray sources & Rate$_{\mathrm{NIR}}$\footnotemark[$*$] & Rate$_{\mathrm{opt}}$\footnotemark[$\dagger$] & Rate$_{\mathrm{dist}}$\footnotemark[$\ddagger$] \\
   \hline
    A           & \phantom{0}89  & 27\% (24)       & \phantom{0}7\% (6)  & \phantom{0}2\% (2) \\
    B           & 189                   & 42\% (80)       & 12\% (23)                 & \phantom{0}8\% (15)\\
    C           & 270                   & \phantom{0}59\% (160)     & 30\% (80)                & 16\% (44) \\
    \hline
     Total     & 548 &  48\% (264) &   20\% (109)  &   11\% (61)\\
     \hline
  \end{tabular}
%\end{center}
\par \noindent
\footnotemark[$*$]NIR Identification rate. The number in parentheses indicates the number of sources.
\footnotemark[$\dagger$]Optical identification rate.
\footnotemark[$\ddagger$]Optical identification rate with a distance estimate by Gaia.
\end{table}

%% file: t06.tex
\setlength{\tabcolsep}{0.2in}
\begin{table*}[htbp]
\small
 \caption{Best-fit parameters for the composite spectra of the  three groups.}\label{t06}
 \begin{tabular}{lcccccc}
  \hline
  \hline
Group  &  \nh\footnotemark[$*$] & \kt$_{1}$\footnotemark[$*$]  & \kt$_{2}$\footnotemark[$*$]   & $Z$\footnotemark[$*$] & $\Gamma$\footnotemark[$*$]  & $\chi^{2}$/d.o.f \\
            &   (10$^{22}$~cm$^{-2}$) & (keV)                                & (keV)         &    &          &                  \\
            \hline
A         & 0.92$_{-0.12}^{+0.14}$    & 4.6$_{-1.9}^{+7.7}$   & --          & 0.21$_{-0.18}^{+0.50}$ &   0.25$_{-0.25}^{+0.60}$&  \phantom{0}55.35/70\\
B         & 0.38$_{-0.12}^{+0.34}$    & 0.7$_{-0.5}^{+0.3}$   &  6.3$_{-1.2}^{+2.0}$   & 1.01$_{-0.87}^{+1.16}$& -- &  \phantom{0}41.14/70\\
C         & 0.40$_{-0.18}^{+0.19}$    & 0.6$_{-0.1}^{+0.3}$   &  2.0$_{-0.2}^{+0.4}$   & 0.22$_{-0.11}^{+0.18}$& -- &  \phantom{0}61.12/70\\
\hline
\end{tabular}
\par \noindent
\footnotemark[$*$]
Uncertainties represent 90\% confidence intervals.
%\footnotemark[$\dagger$]
%Exposure time.
%\footnotemark[$\ddagger$]
% Average seeing of all images in each mask.
\end{table*}

%% file: t07.tex
\begin{table}[htbp]
\small
 \caption{Best-fit parameters for Fe K line fitting.}\label{t07}
%\begin{center}
 \begin{tabular}{ccc}
  \hline
   \hline
   Group  & Energy center\footnotemark[$*$]  & EW\footnotemark[$*$] \\
              & (keV)  &        (eV)                  \\
   \hline
   A    &  6.67$_{-0.11}^{+0.10}$ &  162$_{-26}^{+18}$\\
   B   &   6.76$_{-0.44}^{+0.13}$  &  488$_{-91}^{+50}$\\
   C   &   6.70 (fix)                        &   3 (upper limit)\\
\hline
\end{tabular}
\par \noindent
\footnotemark[$*$]
Uncertainties represent 90\% confidence intervals. 
%\end{center}
\end{table}

%% file: t08.tex
\setlength{\tabcolsep}{0.08in}
\begin{table}[htbp]
\small
 \caption{Time variable sources.}\label{t08}
 %\begin{center}
 \begin{tabular}{ccc}
  \hline
  \hline
  Group  & Definitely\footnotemark[$*$] & Possibly\footnotemark[$*$]\\
  \hline
  A & 0 & 1\\
  B & 5 & 6\\
  C & 6 & 12\\
  \hline
  Sum & 11 & 19 \\
\hline
\end{tabular}
\par \noindent
  %\end{center}
\par \noindent
\footnotemark[$*$] Based on the null hypothesis probability of the K-S test:
 $<5\times10^{-3}$ (definitely variable) or $5\times10^{-3}$--$5\times10^{-2}$
 (possibly variable).
\end{table}

%% file: t09.tex
\setlength{\tabcolsep}{0.08in}
\begin{table}[htbp]
\small
 \caption{Results of NIR spectroscopy.}\label{t09}
%\begin{center}
 \begin{tabular}{cll}
  \hline
   \hline
  Group\footnotemark[$*$] & CBF\footnotemark[$\dagger$] & CPF\footnotemark[$\dagger$] \\
  \hline
  A & H+abs (3) & H+emi (2), H+abs (5)\\
  B & H+abs (7), S+abs (4) & H+abs (15), S+abs (4)\\
  C & S+abs (9) & S+abs (14)\\
  \hline
 \end{tabular}
 \par 
 \noindent
 \footnotemark[$*$] Groups defined in the X-ray color-color diagram (figure~\ref{f5}).
 \footnotemark[$\dagger$] Number of sources in the parentheses based on the X-ray
 spectral hardness (``S''oft or ``H''ard) + NIR ``emi''ssion or ``abs''orption
 features.
 %\end{center}
 %\footnotemark[$*$] Flux of the gaussian component.
\end{table}